\documentclass[12pt]{article}
\usepackage{fullpage,citesort,tableau,epsfig,psfrag,graphics,amsbsy,amssymb}
\usepackage{caption}
\newcommand{\phys}{{\rm phys}}

\newcommand{\beq}{\begin{equation}}
\newcommand{\eeq}{\end{equation}}
\newcommand{\bea}{\begin{eqnarray}}
\newcommand{\eea}{\end{eqnarray}}

\newcommand{\be}{\begin{equation}}
\newcommand{\ee}{\end{equation}}
\newcommand{\bq}{\begin{eqnarray}}
\newcommand{\eq}{\end{eqnarray}}
\newcommand{\ket}[1]{|#1\rangle}
\newcommand{\bra}[1]{\langle#1|}

\newcommand{\eg}{{\it e.g.\ }}

\def\math{\mathsurround=0pt }
\def\leftrightarrowfill{$\math \mathord\leftarrow \mkern-6mu 
 \cleaders\hbox{$\mkern-2mu \mathord- \mkern-2mu$}\hfill
 \mkern-6mu \mathord\rightarrow$}
\def\overleftrightarrow#1{\vbox{\ialign{##\crcr
     \leftrightarrowfill\crcr\noalign{\kern-1pt\nointerlineskip}
     $\hfil\displaystyle{#1}\hfil$\crcr}}}

\newcommand{\VEV}[1]{\langle#1\rangle}

\let\l=\lambda

 \def\bd{\begin{document}} \def\ed{\end{document}}
\def\ds{\documentstyle} \let\fr=\frac \let\bl=\bigl \let\br=\bigr
\let\Br=\Bigr \let\Bl=\Bigl
\let\bm=\bibitem
\let\na=\nabla
\let\pa=\partial \let\ov=\overline
\def\ft#1#2{{\textstyle{{\scriptstyle #1}\over {\scriptstyle #2}}}}
\def\fft#1#2{{#1 \over #2}}
\def\vp{\varphi}
\def\sst#1{{\scriptscriptstyle #1}}
\def\oneone{\rlap 1\mkern4mu{\rm l}}
\def\td{\tilde}
\def\wtd{\widetilde}
\def\dalemb#1#2{{\vbox{\hrule height .#2pt
        \hbox{\vrule width.#2pt height#1pt \kern#1pt
                \vrule width.#2pt}
        \hrule height.#2pt}}}
\def\square{\mathord{\dalemb{6.8}{7}\hbox{\hskip1pt}}}
\def\wtd{\widetilde}
\def\R{\rlap{\rm I}\mkern3mu{\rm R}}
\def\im{{\rm i}}
\def\tilg{\tilde{g}}
\def\tilF{\tilde{F}}
\def\tilA{\tilde{A}}
\def\varf{\varphi}
\def\tilf{\tilde{\phi}}
\def\tilh{\tilde{h}}
\def\rme{{\rm e}}
\def\ep{\epsilon}
\def\0{{(0)}}
\def\9{{(9)}}
\def\8{{(8)}}
\def\7{{(7)}}
\def\6{{(6)}}
\def\5{{(5)}}
\def\4{{(4)}}
\def\3{{(3)}}
\def\2{{(2)}}
\def\1{{(1)}}
\newcommand{\trace}{{\rm Tr}}
\newcommand{\ub}{\overline{U}}
\newcommand{\vb}{\overline{V}}
\newcommand{\uh}{\widehat{U}}
\newcommand{\vh}{\widehat{V}}
\newcommand{\ubh}{\overline{\widehat{U}}}
\newcommand{\vbh}{\overline{\widehat{V}}}
\newcommand{\lb}{\bar{\l}}
\newcommand{\Fb}{\overline{F}}
\newcommand{\Fh}{\widehat{F}}
\newcommand{\Fbh}{\overline{\widehat{F}}}
\newcommand{\Ab}{\overline{A}}
\newcommand{\Ah}{\widehat{A}}
\newcommand{\Abh}{\overline{\widehat{A}}}
\newcommand{\Gb}{\overline{G}}
\newcommand{\Gh}{\widehat{G}}
\newcommand{\Gbh}{\overline{\widehat{G}}}
\newcommand{\Pb}{\overline{P}}
\newcommand{\Ph}{\widehat{P}}
\newcommand{\Pbh}{\overline{\widehat{P}}}
\newcommand{\Qb}{\overline{Q}}
\newcommand{\Qh}{\widehat{Q}}
\newcommand{\Qbh}{\overline{\widehat{Q}}}
\newcommand{\Bb}{\overline{B}}
\newcommand{\Bh}{\widehat{B}}
\newcommand{\Bbh}{\overline{\widehat{B}}}
\newcommand{\fhns}{\hat{F}^{\rm (NS)}}
\newcommand{\fhrr}{\hat{F}^{\rm (RR)}}
\newcommand{\ahns}{\hat{A}^{\rm (NS)}}
\newcommand{\ahrr}{\hat{A}^{\rm (RR)}}
\newcommand{\hhrr}{\hat{H}^{\rm (RR)}}
\newcommand{\hchi}{\hat{\chi}}
\newcommand{\hphi}{\hat{\phi}}
\newcommand{\htau}{\hat{\tau}}
\newcommand{\cG}{{\cal G}}
\newcommand{\cGb}{\overline{{\cal G}}}
\newcommand{\cH}{{\cal H}}
\newcommand{\cP}{{\cal P}}
\newcommand{\cPb}{\overline{{\cal P}}}
\newcommand{\cQ}{{\cal Q}}
\newcommand{\cQb}{\overline{{\cal Q}}}
\newcommand{\cM}{{\cal M}}
\newcommand{\cN}{{\cal N}}
\newcommand{\cO}{{\cal O}}
\newcommand{\cD}{{\cal D}}
\newcommand{\cL}{{\cal L}}
\newcommand\ellhatz{{\hat L}_0}

\newcommand{\vpp}{\mbox{$\langle{\scriptstyle++}\rangle$}}
\newcommand{\vmp}{\mbox{$\langle{\scriptstyle-+}\rangle$}}
\newcommand{\vppp}{\mbox{$\langle{\scriptstyle+++}\rangle$}}
\newcommand{\vmpp}{\mbox{$\langle{\scriptstyle-++}\rangle$}}
\newcommand{\vpmp}{\mbox{$\langle{\scriptstyle+-+}\rangle$}}

\begin{document}
\setlength{\captionmargin}{36pt}
\begin{titlepage}
\phantom{.}

\vskip 3cm
\begin{center}
\begin{large}
{\bf Improved Proof of the No-ghost Theorem\\ for Fermion States
of the Superstring}
\footnote{Supported in part by the Department
of Energy under Grant No. DE-FG02-97ER-41029.} 
\end{large}

\vskip 2cm
{\large 
Charles B. Thorn\footnote{E-mail  address: {\tt thorn@phys.ufl.edu}}
}
\vskip0.8cm
{\it Institute for Fundamental Theory,\\
Department of Physics, University of Florida,
Gainesville FL 32611}


\vskip 1.0cm
\end{center}

\begin{abstract}
\noindent The purpose
of this note is to extend the
improved proof of the no-ghost theorem for the bosonic
and Neveu-Schwarz dual resonance
models, presented in my article Nuclear Physics B{\bf 286} (1987) 61,
to cover the Ramond fermion string. As in that paper, the improvement
involves the identification of an efficient basis for string
state space and a self-contained proof, based on the
super-Virasoro algebra, of the linear independence
of the basis elements. We use our results to calculate the
BRST cohomology for this system.
\end{abstract}
\vfill
\end{titlepage}
\section{Introduction}
The original proof of the no-ghost theorem \cite{goddardt,brower}
for dual resonance models (aka string theory)
was substantially streamlined (and improved) some time ago 
\cite{thorndetailed}. The latter work focussed on the bosonic
open string model, but it also sketched straightforward extensions to
the Neveu-Schwarz model \cite{neveuschwarz} and to the associated closed
string models. However, the extension of the improved
proof to the fermion (Ramond) sector \cite{ramond,neveustfermions} 
of the superstring model \cite{gliozziso} was not included
because of complications due to the presence of fermionic zero modes.
The purpose of this short note is to fill this lacuna. As a
bonus, we also apply our results to the calculation of
the BRST cohomology of the Ramond sector of the superstring.

The original arguments, especially those in \cite{goddardt}, 
relied on an efficient basis of the state
space that was an extension of one proposed in an earlier
paper \cite{browert}. 
In particular, the linear independence of these basis states 
was proved in a way that, as in \cite{browert},
relied on the explicit representation of the (super)-Virasoro
operators in terms of normal mode oscillators. 
The improved arguments,
offered in \cite{thorndetailed}, established the required 
linear independence using only the algebra of the operators
without regard to their representation. However, once the linear independence
was established, the remainder of the proof followed exactly as in 
\cite{goddardt}. The improved proof was completely self-contained
and significantly more efficient than the originals.

In the Ramond sector states generated by the super-Virasoro
operators $L_n,F_n$, namely states of the form 
$L_{-n}\ket{\psi_1}+F_{-n}\ket{\psi_2}$ for $n>0$, decouple
from physical amplitudes. Physical states are therefore
defined as the orthogonal complement to the space of these
decoupled states, in other words,
\bea
L_n\ket{{\rm phys}} = F_n\ket{{\rm phys}} =0,\qquad {\rm for}\quad n>0
\eea
It is convenient to call such states physical even 
when they do not also satisfy the on-mass-shell condition
$F_0\ket{{\rm phys}}=0$.

The super-Virasoro algebra in $D$ spacetime dimensions reads:
\bea
 {}[L_n, L_m]&=&(n-m)L_{n+m}+{D\over8}n^3\delta_{n,-m}\\
 {}[L_n, F_m]&=&\left({n\over2}-m\right)F_{n+m}\\
 {}\{F_n, F_m\}&=&2L_{n+m}+{D\over2}n^2\delta_{n,-m}\;.
\eea
The special features that emerge in the critical dimension
$D=10$ figure prominently in both the original
proof and the improved versions.

As long as the spacetime dimension $D\geq2$, there is a light-like
direction $k^\mu$ such that $k\cdot p\neq0$, where $p^\mu$
is the energy-momentum eigenvalue of the string state.
Then one can introduce the operators $D_n=k\cdot d_n$ and
$K_n=k\cdot a_n$ with $k$ normalized so that $K_0=1$.
Here $a_n^\mu$ are the (bosonic)
normal mode oscillators for the
coordinate $x^\mu(\sigma,\tau)$ and $d_n^\mu$ are the (fermionic)
normal modes of
the worldsheet Dirac operators $\Gamma^\mu(\sigma,\tau)$. It will
be important in what follows that the (anti)commutators of the
operators $K_n,D_n$ among each other all vanish.
We will also require their algebra with the super-Virasoro
operators:
\bea
 {}[L_n,K_m]&=&-mK_{m+n},\qquad [L_n,D_m]=-\left(m+{n\over2}\right)D_{m+n}\\
 {}[F_n,K_m]&=&-mD_{m+n},\qquad \{F_n,D_m\}=K_{m+n}\\
 {}[K_n,K_m]&=&0,\qquad\{D_n,D_m\}=0,\qquad [K_n,D_m]=0\;.
\eea
The physical states that, in addition, satisfy $K_n\ket{{\rm phys}}=0$
for $n>0$ and $D_n\ket{{\rm phys}}=0$ for $n\geq0$ 
are called the transverse states \cite{ddf}. It follows from
$\{D_0,F_0\}=K_0=1$ that any two transverse states have vanishing
inner product:
\bea
\VEV{T|T^\prime}&=&\bra{T}\{D_0,F_0\}\ket{T^\prime}
=\bra{T}(D_0F_0+F_0D_0)\ket{T^\prime}=0.
\eea
Here it is understood that $\bra{T}$ is the appropriate Dirac adjoint
under which the zero mode $d_0^\mu$, which can be represented as
proportional to a Dirac matrix $\gamma^\mu$, is self adjoint.
With this understanding, nonzero inner products require the insertion of
an $F_0$ factor:
\bea
\bra{T}F_0\ket{T^\prime}&\neq&0\; .
\eea
Defining the norm with this inner product, the transverse states
have nonnegative norm, relative to an overall constant factor. 
To see this note that they belong to the
larger space of states $\ket{\phi}$ 
generated by $D_{-k},K_{-k}, a^i_{-k}, d^i_{-k}$ where
$1\leq i\leq D-2$ acting on a Fock vacuum $\ket{0}$\footnote{
By virtue of the zero modes $d_0^\mu$, the lowest mass
level is degenerate. We can single one out to use as $\ket{0}$
by, for example, forming the combinations $d_0^k\pm i d_0^{k+1}$
for $k=1,3,5,7$ and defining $\ket{0}$ to satisfy
\bea
(d_0^k+id_0^{k+1})\ket{0}&=&D_0\ket{0}=0,\qquad{\rm and}\quad d_n^\mu\ket{0}
=a_n^\mu\ket{0}=0,
\quad {\rm for}~n>0. 
\eea
It is, of course, 
understood that $\ket{0}$ has energy momentum eigenvalue $p^\mu$.}. 
Then the
inner product
\bea
{\bra{\phi}F_0\ket{\phi^\prime}\over
\bra{0}p\cdot d_0\ket{0}}&\geq0\; .
\label{inner}
\eea 
Then a basis for the whole space can
be taken of the form \cite{goddardt,thorndetailed}
\bea
\ket{\{f\}\{\lambda\},\{d\}\{\kappa\}}
=F_0^{f_0}F_{-1}^{f_1}L_{-1}^{\lambda_1}\cdots 
F_{-l}^{f_l}L_{-l}^{\lambda_l}
D_{-1}^{d_1}\cdots D_{-k}^{d_k}
K_{-1}^{\kappa_1}\cdots K_{-k}^{\kappa_k}\ket{T}
\label{lkbasis}
\eea
where $\ket{T}$ are arbitrary transverse states.
Also $\{\lambda\}$ and $\{\kappa\}$ are bosonic partitions of 
two nonnegative integers. Similarly $\{f\}$ and $\{d\}$
are fermionic partitions of two nonnegative integers. 
Fermionic simply means that each $f_i$ and $d_i$
assumes only the values 0 or 1. 

\section{Ordering the Basis}
The basis (\ref{lkbasis}) is labeled by partitions of  
mode number. Partitions of an integer $N$ are labeled by
a sequence of nonnegative integers $p_1,p_2,\cdots,p_N$
such that $N=\sum_{i=1}^N ip_i$. For bosonic partitions
the $p_i$ can assume any nonnegative integer value, whereas
for fermionic partitions each $p_i$ is either $0$ or $1$.
A natural way to order partitions is ``lexicographically''
or ``alphabetically''. This means that $\{p\}<\{p^\prime\}$
if the first nonzero entry of the sequence
\bea
\sum_i i(p_i-p_i^\prime),\quad p_1-p_1^\prime,\quad p_2-p_2^\prime,\quad
\cdots
\eea
is positive. 

The basis we employ in this paper
is actually labeled by four partitions $\{f\},\{d\},\{\lambda\},\{\kappa\}$.
The first two are fermionic and the last two are bosonic. The super-Virasoro
generators are controlled by $\{f\},\{\lambda\}$, and it is convenient
to order them jointly according to $(\{f,\lambda\})<(\{f^\prime,
\lambda^\prime\})$ if the first nonzero entry of the sequence
\bea
\sum_{i}i(f_i-f_i^\prime+\lambda_i-\lambda_i^\prime),\quad f_0-f_0^\prime, 
\quad f_1-f_1^\prime,\quad \lambda_1-\lambda_1^\prime,\quad f_2-f_2^\prime,
\quad
\cdots
\eea
is positive. The operators $K_n=k\cdot a_n$ and $D_n=k\cdot d_n$
are controlled by the other pair of partitions $\{d\kappa\}$.
Then the basis elements are labeled by $(\{f\lambda\},\{d\kappa\})$.
Then we order the entire basis according to $(\{f\lambda\},\{d\kappa\})
<(\{f^\prime\lambda^\prime\},\{d^\prime\kappa^\prime\})$
if $\{f\lambda\}<\{f^\prime\lambda^\prime\}$ or if 
$\{f\lambda\}=\{f^\prime\lambda^\prime\}$ and
$\{d\kappa\}>\{d^\prime\kappa^\prime\}$.
\section{Linear Independence}
To prove linear independence of the basis elements, it is
convenient to define a conjugate (or ``dual'') to each
element (\ref{lkbasis}) as follows
\bea
\ket{\{f\}\{\lambda\},\{d\}\{\kappa\},C}
=F_0^{1-f_0}F_{-1}^{d_1}L_{-1}^{\kappa_1}\cdots 
F_{-k}^{d_k}L_{-k}^{\kappa_k}
D_{-1}^{f_1}\cdots D_{-l}^{f_l}
K_{-1}^{\lambda_1}\cdots K_{-l}^{\lambda_l}\ket{T}\; .
\label{conjbasis}
\eea
As we shall see shortly the inner product of each basis element
with its conjugate is not zero.

But our first task is to prove that
\bea
\VEV{\{f\}\{\lambda\},\{d\}\{\kappa\},C|\{f^\prime\}\{\lambda^\prime\},
\{d^\prime\}\{\kappa^\prime\}}=0,\qquad {\rm if}\quad(\{f\lambda\},\{d\kappa\})<
(\{f^\prime\lambda^\prime\},\{d^\prime\kappa^\prime\})\; ,
\eea
which is to say that the corresponding matrix of inner products
is lower triangular. The proof is a recursive one in which we
consider in turn the ways in which $(\{f\lambda\},\{d\kappa\})<
(\{f^\prime\lambda^\prime\},\{d^\prime\kappa^\prime\})$. We organize
the argument as a series of steps.
\begin{description}
\item{\bf Step~1.} We first suppose 
$\sum_{i}i(f_i-f_i^\prime+\lambda_i-\lambda_i^\prime)>0$. Then the
process of moving the positive moded $F_n,L_n$ to the right, whence
they annihilate $\ket{T}$, leaves behind a state with negative moded
$F_{-n},L_{-n}$ with an even smaller total mode number. Moving these to the
left, whence they annihilate $\bra{T}$, leaves behind a matrix element
with at least one $K_n$ or $D_n$ with non-zero mode number. 
Since $k$ is light-like
all these operators (anti)commute with each other and any one of them
kills the matrix element, establishing the claim.
\item{\bf Step~2.} So next suppose $\sum_{i}i(f_i-f_i^\prime+\lambda_i-\lambda_i^\prime)=0$ and $f_0>f_0^\prime$, or $f_0=1,f_0^\prime=0$ since $\{f\}$
is a fermionic partition. In this case we see that all factors of
$F_0$ are initially absent from the matrix element. 
After the matrix element is
completely reduced, it will be nonzero only if a single factor
of $F_0$ is left behind. Since under the conditions of this step,
no $F_0$ is initially present, such a factor must come from a commutator
such as $[L_n,F_{-n}]$ or $[F_n,L_{-n}]$ for some nonzero $n$ 
when e.g. the positive moded
super-Virasoro operators are moved to the right. But this would
leave behind negative moded super-Virasoro operators of smaller
total mode number than we started with, and the remaining evaluation
would meet the conditions of step 1, establishing the claim.
\item{\bf Step~3.} Now we suppose 
$\sum_{i}i(f_i-f_i^\prime+\lambda_i-\lambda_i^\prime)
=f_0-f_0^\prime=0$ and $f_1>f_1^\prime$,
or $f_1=1,f_1^\prime=0$ since $\{f\}$ is a fermionic partition.
This means the bra contains one $D_1$ and the ket contains
no $F_{-1}$. Also there is precisely one factor of $F_0$
present. Now move the $D_1$ operator to the right until
it annihilates $\ket{T}$. An (anti)commutator of $D_1$ with
a positive moded ($F_n$)$L_n$ produces a $(K_{n+1})D_{n+1}$,
both of which increase the mode number of the $K_n,D_n$
factors in the bra producing the conditions of Step 1 and
so yield a vanishing contribution. An anticommutator with
$F_0$ replaces $F_0$ with $K_1$ creating the conditions
of Step 2 and a vanishing contribution. There is no
$F_{-1}$ in the ket under the conditions of this step, so
the next possibility is a commutator with $L_{-1}$ which produces
a $D_0$. Moving this $D_0$ to the right until it annihilates
$\ket{T}$ picks up (anti)commutators with negative moded
$(F_{-n})L_{-n}$ which reduce the total mode number of 
the super-Virasoro operators in the ket compared to the $K,D$
in the bra producing the conditions of Step 1 and a vanishing
contribution. 
Finally an (anti)commutator of $D_1$ with $(F_{-n})L_{-n}$ with $n>1$
produces a $(D_{-(n-1)})K_{-(n-1)}$ both of which are negative moded.
This procedure reduces the mode number of the super-Virasoro
operators in the ket by $n>1$ whereas it reduces the mode number of
$D_n,K_n$ in the bra by only 1, creating the conditions of
Step 1 and a vanishing contribution. Thus the claim is established.
\item{\bf Step~4.} Now suppose $\sum_{i}i(f_i-f_i^\prime+\lambda_i-\lambda_i^\prime)=f_0-f_0^\prime=f_1-f_1^\prime=0$ and $\lambda_1>\lambda_1^\prime$.   
since $\{\lambda\}$ is bosonic $\lambda_1^\prime$ is allowed to
be nonzero. There is precisely one $D_1$ in the bra and one $F_{-1}$
in the ket. First move $D_1$ to the right. From the reasoning of
Step 3 the only nonvanishing contribution comes from the
anticommutator with $F_{-1}$ which produces $K_0$ a nonzero number.
After this reduction the matrix element contains neither $D_1$
nor $F_{-1}$, and we turn to reducing the $K_1$'s by moving them
to the right. Bearing in mind the considerations in Step 3,
we see that the only nonvanishing contributions come from the
commutator $[K_1,L_{-1}]$ which produces a $K_0$. Since $\lambda_1
>\lambda_1^\prime$, there will still be left over $K_1$'s in the
bra after all the $L_{-1}$'s are removed from the ket. Moving
them to the right produces no nonvanishing contributions,
establishing the claim.
\item{\bf Step~5.} Now suppose 
$\sum_{i}i(f_i-f_i^\prime+\lambda_i-\lambda_i^\prime)
=f_0-f_0^\prime=f_1-f_1^\prime=\lambda_1-\lambda_1^\prime=0$ 
and $f_2>f_2^\prime$. Start by removing the $D_1,K_1,F_{-1},L_{-1}$
factors from the matrix element. Then moving $D_2$ to the right
can give no nonvanishing contributions since $f_2=1, f_2^\prime=0$.
Proceeding in this way we eventually see that the matrix element is
zero unless $\{f\lambda\}=\{f^\prime\lambda^\prime\}$.
\item{\bf Step 6.} Finally we suppose $\{f\lambda\}=\{f^\prime\lambda^\prime\}$
and $\{d\kappa\}>\{d^\prime\kappa^\prime\}$. Start
 by reducing out all the $K_n,D_n$ in the bra against all the $L_{-n},
F_{-n}$ in the ket. This leaves the $L_n,F_n$ in the bra
and the $D_{-n},K_{-n}$ in the ket. A fortiori the mode numbers
are equal and there is precisely one $F_0$ in the matrix element.
So now one repeats Steps 3-5 with the role of bra and ket interchanged
and $\{d^\prime\kappa^\prime\},\{d\kappa\}$ playing the
roles of $\{f\lambda\},\{f^\prime\lambda^\prime\}$ respectively.
The claim is then established.
\end{description}
As a corollary to the detailed considerations of the above argument it
follows that
\bea
\VEV{\{f\}\{\lambda\},\{d\}\{\kappa\},C|\{f\}\{\lambda\},
\{d\}\{\kappa\}}\neq0,
\eea
In particular the transverse space is positive definite under the
norm defined by the inner product (\ref{inner}).
These conclusions follow because the only nonzero contributions 
that arise in the reduction 
process come from one of the (anti)commutators
$$[K_n,L_{-n}],\quad \{D_{n},F_{-n}\},\quad [L_n,K_{-n}],\quad 
\{F_{n},D_{-n}\},$$
all of which are proportional to $K_0$ a nonzero number.

We have therefore established that the matrix 
\bea
\VEV{\{f\}\{\lambda\},\{d\}\{\kappa\},C|\{f^\prime\}\{\lambda^\prime\},
\{d^\prime\}\{\kappa^\prime\}}
\eea
is triangular with nonzero diagonal entries. It follows that
its determinant is not equal to zero, and hence that
the states (\ref{lkbasis}) are linearly independent. In other words
these states form a basis of the whole state space.
\section{Proof of No Ghost Theorem}
Armed with the linearly independent basis (\ref{lkbasis}) 
the proof of the no ghost theorem in the Ramond sector follows the
original one \cite{corrigang}.
The condition on on-shell physical states is that they are annihilated by
$F_n,L_n$ for $n\geq0$. Because of the super-Virasoro algebra, it is
sufficient to impose only the two conditions
\bea
F_0\ket{\rm phys}&=&0,\qquad L_1\ket{\rm phys}=0\; .
\eea
Using the basis defined in the previous section, the first condition
implies that
\bea
\ket{\rm phys}&=& F_0\ket{\psi}
\eea
for some $\ket{\psi}$. This ket can be expanded in terms of the basis elements
that have no $F_0$ factor. It can be decomposed as
\bea
\ket{\psi}&=&\ket{s}+\ket{\phi}\; ,
\eea
where $\ket{\phi}$ contains only the basis elements that have {\it no}
super-Virasoro factors (i.e. $\{f\}=\{\lambda\}=0$, and $\ket{s}$
contains basis elements with at least one negative moded super-Virasoro
generator. Since $L_{-1},F_{-1}$ generate via the algebra all other
$F_{-n},L_{-n}$, it follows that we can write
\bea
\ket{s}&=&L_{-1}\ket{\psi_1}+F_{-1}\ket{\psi_2}\; .
\eea
We now compute the action of $L_1$:
\bea
L_1F_0(L_{-1}\ket{\psi_1}+F_{-1}\ket{\psi_2})&=&
F_0(L_{-1}L_1\ket{\psi_1}+F_{-1}L_1\ket{\psi_2})
+{1\over2}F_1(L_{-1}\ket{\psi_1}+F_{-1}\ket{\psi_2})\nonumber\\
&&+F_0\left[\left(2L_0+{D\over8}\right)\ket{\psi_1}
+{3\over2}F_0\ket{\psi_2}\right]\nonumber\\
&=&F_0\ket{s^\prime}+\ket{s^{\prime\prime}}
+{3\over4}F_0\ket{\psi_1}+\left(L_0+{D\over4}\right)\ket{\psi_2}\nonumber\\
&&+F_0\left(2L_0+{D\over8}\right)\ket{\psi_1}
+{3\over2}L_0\ket{\psi_2}\nonumber\\
&=&F_0\ket{s^\prime}+\ket{s^{\prime\prime}}
+\left({D\over4}-{5\over2}\right)
\ket{\psi_2}+F_0\left({D\over8}-{5\over4}\right)\ket{\psi_1}\nonumber\\
&\to&F_0\ket{s^\prime}+\ket{s^{\prime\prime}},\qquad {\rm for}~D=10\\
L_1F_0\ket{\phi}&=&F_0\ket{\phi^\prime}+\ket{\phi^{\prime\prime}}\; .
\eea
The key point is that, in the critical dimension $D=10$,
the physical state condition
\bea
L_1F_0(\ket{s}+\ket{\phi})=0\qquad {\rm implies}
\quad L_1F_0\ket{s}=L_1F_0\ket{\phi}=0.
\eea
But $L_1F_0\ket{s}=0$ means $F_0\ket{s}$ is a null spurious state, and
the $L_1F_0\ket{\phi}=0$ implies that $\ket{\phi}=\ket{T}$. Hence
\bea
\ket{\rm phys}&=&F_0\ket{T}+\ket{\rm Null}\; ,\qquad {\rm for}\quad D=10,
\eea
which establishes the no-ghost theorem in the fermion sector of the
superstring.
\section{BRST Cohomology}
Just as was done in \cite{thorndetailed} for the bosonic string, 
the basis (\ref{lkbasis}) can be used to give an efficient
calculation of the cohomology of the BRST operator for the Ramond
sector of the superstring. For other approaches to
BRST cohomology see \cite{katoogawa}. The argument below parallels that in
\cite{thorndetailed} quite closely, and indeed was sketched
in my review of string field theory \cite{thornsft} for both
the Neveu-Schwarz and Ramond sectors. Here, for the sake
of completeness we give a more detailed 
calculation for the Ramond sector.

Recall that the BRST method starts with the construction of
a nilpotent Grassmann odd operator
\bea
Q &=& \sum_m c_{-m}L_m +\sum_m\gamma_{-m}F_m -
{1\over 2}\sum_{m,n}(m-n):c_{-m}c_{-n}b_{m+n}: \nonumber\\
&&\quad -\sum_{m,n}
\gamma_{-m}\gamma_{-n} b_{m+n}+
\sum_{m,n}({3m\over 2}-n):c_{-m}\gamma_{m-n}\beta_n:\\
Q^2&=&0\qquad {\rm for}~D=10
\label{ramondbrst}
\eea
for the Ramond sector of the open superstring. Here $b,c$ are the
fermionic
reparameterization ghosts and $\beta,\gamma$ are the bosonic superghosts.
\bea
\{c_m,b_n\} = \delta_{m,-n},\qquad [\gamma_m,\beta_n]=\delta_{m,-n}
\eea
all other graded brackets vanishing. Total ghost number is defined
as
\bea
G&=&G^b+G^\beta = c_0b_0 + \sum_{k\neq0}:c_{-k}b_k: 
+\gamma_0\beta_0- \sum_{k\neq0}:\gamma_{-k}\beta_k: 
\eea
where the colons denote normal ordering in the usual way.

In the BRST formalism physical states are identified with the
cohomology of $Q$:
\bea
Q\ket{\phys}=0, \qquad \ket{\phys}\equiv \ket{\phys}+Q\ket{\Lambda}
\label{openfieldeq}
\eea
Here we will use the basis (\ref{lkbasis}) augmented by ghost 
excitations to calculate the cohomology. The argument is
significantly eased by employing the following replacements for
$F,L,c,\gamma$:
\bea
F_n&\to& {\hat F}_n\equiv [Q,\beta_n]\\
L_n&\to& {\hat L}_n\equiv \{Q,b_n\}\\
c_n&\to& {\hat c}_n\equiv [Q,K_n],\qquad n\neq0\\
\gamma_n&\to& {\hat \gamma}_n\equiv \{Q,D_n\}.
\eea
Note that since $[Q,K_0]=0$, we 
retain $c_0$ in its unmodified form.
Except for $c_0$, these ``dressed'' operators have zero 
graded brackets with $Q$.
Furthermore all the operators ${\hat c}_n,{\hat\gamma}_n,K_n,D_n$
retain vanishing brackets with each other. In
addition, the ${\hat F}_m,{\hat L_n}$ satisfy the 
super-Virasoro algebra with no c-number term.

We shall identify the transverse states $\ket{T}$ in the basis
(\ref{lkbasis}) with the states in the larger BRST space
that satisfy
\bea
(b_0,D_0,\beta_0,b_n,{\hat c}_n,\beta_n,{\hat\gamma}_n,K_n,{\hat L}_n,
{\hat F}_n,D_n)\ket{T}=0,\quad n>0.
\label{brsttransverse}
\eea
which are equivalent to the same conditions with the hats removed.
Choosing the transverse states to be 
annihilated by $\beta_0$ amounts to a choice
of picture (see for example \cite{thornsft}) and assigns 0 total ghost number
to the $\ket{T}$.
Then a basis for the larger space is obtained by applying
to (\ref{lkbasis}) independent monomials in ${\hat c}_{-n},{\hat\gamma}_{-n}$
with $n\geq0$, and independent monomials in $b_{-n},\beta_{-n}$ for
$n>0$.

A final efficiency is gained by changing from the ``occupation
number'' labelling of (\ref{lkbasis}) to (an)a (anti)symmetric
tensor labelling:
\bea
{\hat F}_{-1}^{f_1}{\hat L}_{-1}^{\lambda_1}\cdots 
{\hat F}_{-l}^{f_l}{\hat L}_{-l}^{\lambda_l}&\leftrightarrow&
{\hat F}_{-[n_1}{\hat F}_{-n_2}
\cdots {\hat F}_{-n_f]}{\hat L}_{-\{m_1}{\hat L}_{-m_2}
\cdots {\hat L}_{-m_l\}}\\
D_{-1}^{d_1}\cdots D_{-k}^{d_k}&\leftrightarrow&{D}_{-[n_1}{D}_{-n_2}
\cdots {D}_{-n_d]}\\
K_{-1}^{\kappa_1}\cdots K_{-k}^{\kappa_k}&\leftrightarrow&
{K}_{-\{n_1}{ K}_{-n_2}
\cdots {K}_{-n_k\}}\\
b_{-1}^{b_1}\cdots b_{-k}^{b_k}&\leftrightarrow&{b}_{-[n_1}{b}_{-n_2}
\cdots {b}_{-n_b]}\\
\beta_{-1}^{\beta_1}\cdots \beta_{-k}^{\beta_k}&\leftrightarrow&
{\beta}_{-\{n_1}{\beta}_{-n_2}
\cdots {\beta}_{-n_\beta\}}\\
c_{-1}^{c_1}\cdots c_{-k}^{c_k}&\leftrightarrow&{c}_{-[n_1}{c}_{-n_2}
\cdots {c}_{-n_c]}\\
\gamma_{-1}^{\gamma_1}\cdots \gamma_{-k}^{\gamma_k}&\leftrightarrow&
{\gamma}_{-\{n_1}{\gamma}_{-n_2}
\cdots {\gamma}_{-n_\gamma\}}
\eea
where $\{\ \}$, $[\ ]$ denote respectively
complete symmetrization,
complete antisymmetrization of the enclosed indices.
The equivalence of these two labelling schemes is simply the known equivalence
between occupation number and (anti)symmetric wave function representations
of states of identical fermions or bosons.
We then arrange the basis elements as follows:
\bea
&&{\hat F}_0^{J}{\hat F}_{-[j_1}
\cdots {\hat F}_{-j_f]}{\beta}_{-\{k_1}
\cdots {\beta}_{-k_\beta\}}
{\hat L}_{-\{l_1}
\cdots {\hat L}_{-l_l\}}{b}_{-[m_1}
\cdots {b}_{-m_b]}\nonumber\\
&&\qquad\qquad {\hat c}_{-[n_1}
\cdots {\hat c}_{-n_c]}{K}_{-\{p_1}
\cdots {K}_{-p_k\}}{\hat \gamma}_{-\{q_1}
\cdots {\hat \gamma}_{-q_\gamma\}}{D}_{-[r_1}
\cdots {D}_{-r_d]}c_0^M{\hat \gamma}_0^N\ket{T}
\label{bigbasis}
\eea
As we show below the method of \cite{thorndetailed} applied to the
Ramond sector, as sketched in \cite{thornsft}, establishes that
the cohomology of $Q$ is a subspace of the space spanned by the
zero mode elements of (\ref{bigbasis}):
\bea
F_0^Jc_0^M{\hat \gamma}_0^N\ket{T}, \qquad N\geq0,\quad J=0,1,\quad M=0,1
\label{zerobasis}
\eea
This dramatic reduction applies without regard to the on-shell condition
$\ellhatz\ket{\phys}=0$. Writing $\ket{\phys}$ as a linear
combination of (\ref{zerobasis}) it is only a matter of
minor algebra, given in \cite{thornsft}, to show that the kernel of
$Q$ is limited to the two states
\bea
\ket{\phys}_0&=&F_0\ket{T_1},\qquad {\rm provided}~\ellhatz\ket{\phys}=0\\
\ket{\phys}_1&=&(\gamma_0+c_0F_0)\ket{T_2}
\eea 
where the subscripts indicate ghost number. Note $Q\ket{\phys}_1=0$
both on and off shell. Off shell we can write
\bea
\ket{\phys}_1&=&Q{b_0\over\ellhatz}\ket{\phys}_1
\eea 
showing that it is trivial. Since 
\bea
\hat L_0 = {p^2\over 2} +\sum_{m=1}^{\infty}(\alpha_{-m}\cdot\alpha_{m}
+d_{-m}\cdot d_m)
+\sum_{m=1}^{\infty}m(c_{-m}b_m+b_{-m}c_m-\gamma_{-m}\beta_m
+\beta_{-m}\gamma_m)\label{openellzero}
\eea
has a continuous spectrum in the neighborhood of $0$ (since $p^2$
is continuous), we can regard the onshell limit of $\ket{\phys}_1$
also as trivial. On the other hand $Q\ket{\phys}_0=0$ only on
shell, so the space $\{\ket{\phys}_0\}$ is the true cohomology. 
This will complete the
BRST demonstration of the no-ghost theorem for the Ramond sector
of the superstring.

The key to proving the basis reduction (\ref{zerobasis})
 is the tableau identity
(see \eg\  
\cite{hamermesh}, Section 7-12) 
\bea
 \underbrace{\tableau{6}}_\lambda\quad\otimes\quad
\Biggl.\tableau{1 1 1 1 1 1}\Biggr\}\beta=(\beta+1)
\Biggl\{\overbrace{\tableau{6 1 1 1 1 1 1}}^{\lambda}\Biggr.\quad+\beta
\Biggl\{\overbrace{\tableau{7 1 1 1 1 1}}^{\lambda+1}\Biggr. 
\label{tableauii} 
\eea
The product of a symmetric tensor with an
antisymmetric tensor produces precisely two symmetry
patterns. The operators appearing in the basis can be
paired off according to how they behave under bracketing
with $Q$. Each pair has one bosonic and one fermionic member.
For  the $R$ sector the pairings are
\bea
\hat F_n &\leftrightarrow& \beta_n,\qquad \hat L_n \leftrightarrow b_n\qquad
K_n \leftrightarrow {\hat c}_n,\qquad D_n \leftrightarrow {\hat \gamma}_n
\label{coordinatepairs}
\eea
The bracket of $Q$ with a product of 
two monomials, one containing
the bosonic members and the other containing the fermionic
members of a pairing, then shows that one of the two
symmetry patterns can be gauged away, and that the requirement
$Q\ket{\phys}=0$ implies the other pattern is absent
in $\ket{\phys}$. To analyze the
cohomology of $Q$ this procedure is applied sequentially to
each pair.

First we compute the brackets
\bea
&&\hskip-1in \left[Q,{\hat F}_0^{J}{\hat F}_{-[j_1}
\cdots {\hat F}_{-j_{f-1}]}{\beta}_{-\{k_1}
\cdots {\beta}_{-k_{\beta+1}\}}
\right]_{\pm}\nonumber\\
&=&(-)^{f+J-1}{\hat F}_0^{J}\sum_l{\hat F}_{-[j_1}
\cdots {\hat F}_{-j_{f-1}]}{\beta}_{-\{k_1}\cdots{\hat F}_{-k_l}
\cdots {\beta}_{-k_{\beta+1}\}}\\
&=&(-)^{f+J-1}{\hat F}_0^{J}\sum_l{\hat F}_{-[j_1}
\cdots {\hat F}_{-j_{f-1}}{\hat F}_{-k_l]}{\beta}_{-\{k_1}\cdots
\VEV{{\beta}_{-k_l}}
\cdots {\beta}_{-k_{\beta+1}\}}\nonumber\\
&&+\ \hbox{other terms with
\ less \ than}\ J+f\ {\hat F}{\rm's}.
\label{fbetaqbracket} \\
 &&\hskip-1in\left[Q,\hat L_{-\{l_1}\hat L_{-l_2}\right.\left.
\cdots\hat L_{-l_{\ell-1}\}}
b_{- [m_1}b_{-m_2}\cdots b_{-m_{b+1}]}\right]_{\pm}\nonumber\\
&=&\sum_k(-)^{k-1}\hat L_{-\{l_1}\cdots\hat L_{-l_{\ell-1}}\
\hat L_{-m_k\}}b_{-[b_1}\cdots\VEV{b_{-m_k}}\cdots b_{-m_{b+1}]}\nonumber\\
&&+\hbox{other terms with\ less\ than}\
 \ell\ \hat L{\rm's}.
\label{lbqbracket} \\
 &&\hskip-1in\left[Q,{\hat c}_{-[n_1}
\cdots {\hat c}_{-n_{c-1}]}{K}_{-\{p_1}
\cdots {K}_{-p_{k+1}\}}\right]_{\pm}\nonumber\\
&=&\sum_k \hat c_{-[n_1}\cdots\hat c_{-n_{c-1}}\
\hat c_{-p_k]\}}K_{-\{p_1}\cdots\VEV{K_{-p_k}}\cdots K_{-p_{k+1}\}}
\label{ckqbracket}\\
 &&\hskip-1in\left[Q,{\hat \gamma}_{-\{q_1}
\cdots {\hat \gamma}_{-q_{\gamma-1}\}}{D}_{-[r_1}
\cdots {D}_{-p_{d+1}\}}\right]_{\pm}\nonumber\\
&=&\sum_k (-)^{k-1}\hat \gamma_{-\{q_1}\cdots\hat \gamma_{-q_{\gamma-1}}\
\hat \gamma_{-r_k\}}D_{-[r_1}\cdots\VEV{D_{-r_k}}\cdots D_{-r_{d+1}\}}
\label{gammadqbracket} 
\eea
where $\VEV{}$ around an operator means it is absent. The ``other terms'' 
on the right of the first two bracket equations 
come from the nonzero commutators picked up in symmetrizing
the first term on the right. There are no such terms for the
last two bracket equations.\footnote{The reordering terms on
the right of the first bracket equation (\ref{fbetaqbracket}) 
introduce $b$ and ${\hat L}$ 
operators not present
on the left. In contrast the right sides of the last three
bracket equations involve only the operators appearing on the left.
We put the ${\hat F}\beta$
factors on the left in (\ref{bigbasis}) so that at no point,
in the sequential reduction to follow, does the reordering process
introduce operators that had previously been eliminated.}

We now begin the sequential reduction of the basis
elements needed to calculate the cohomology of $Q$, starting with the
${\hat F}\beta$ factors.
 The sum on the right side of (\ref{fbetaqbracket})  
can be recognized as the
Young symmetrizer for the tableaux
\bea f\Biggl\{\overbrace{\tableau{7 1 1 1 1 1}}^{\beta+1}\Biggr..
\label{tableaui}
\eea
If we now apply (\ref{fbetaqbracket}) to a state in the subspace ${\cal K}$
spanned
by the states in (\ref{bigbasis}) with no ${\hat F}_{-(n>0)}$'s 
or $\beta$'s, we learn that
the linear combinations of states in (\ref{bigbasis}) with $f$
${\hat F}_{-(n>0)}$'s and $\beta$ $\beta_{-n}$'s 
with the symmetry (\ref{tableaui}) can be expressed
as a pure gauge plus states with less than $f$
${\hat F}_{-(n>0)}$'s. 
Since we can work in an eigenspace
of $\ellhatz$  there is a maximum possible
value of $f$. Working recursively downward in
$f$ we can systematically remove all linear combinations
with symmetry (\ref{tableaui}).
Then the tableaux identity (\ref{tableauii})
implies that the basis elements (\ref{bigbasis}) can be restricted
to linear combinations for which terms with $f$ ${\hat F}_{-(n>0)}$'s
and $\beta$ $\beta$'s are restricted to the symmetry pattern
\bea
 f+1\Biggl
\{\overbrace{\tableau{6 1 1 1 1 1 1}}^{\beta}
\Biggr..
\label{tableauiii} 
\eea
Now consider (\ref{openfieldeq}) with $|\phys\rangle$ a linear
combination of the states (\ref{bigbasis}) where the $\ell$'s and $b$'s
are symmetrized according to (\ref{tableauiii}) 
and use (\ref{fbetaqbracket}) to conclude
that
\bea
&&\hskip-.5inQ\sum_{\{j,k\}}\left({\hat F}^f_{-\{j\}}
\beta^\beta_{-\{k\}}\right)
_{ \{\beta,1^{f}\}}
\Bigg|\phi,\{\beta,1^{f}\}
\Bigg\rangle\nonumber\\
&=&
\sum\left[\beta
\left({\hat F}^{f+1}_{-\{j\}}\beta^{\beta-1}_{-\{k\}}\right)
_{\{\beta,1^{f}\}}+\ \hbox{terms with
\ less \ than}\ f+1\ {\hat F}{\rm's}\right]
\Bigg|\phi, \{\beta,1^{f}\}
\Bigg\rangle\nonumber\\
&&\hskip1in+\sum
\left({\hat F}^{f}_{-\{j\}}\beta^{\beta}_{-\{k\}}\right)
_{\{\beta,1^{f}\}}
\Bigg|\phi',{-\{j\}},{-\{k\}}
\Bigg\rangle=0
\label{firstfieldeq}
\eea
where the superscripts indicate the number of $\hat F$'s and
$\beta$'s in the polynomials in parentheses.
Also $|\phi,\{\beta,1^{f}\} \rangle$,
$|\phi',{-\{j\}},{-\{k\}} \rangle$ belong to
${\cal K}$. In (\ref{firstfieldeq}),
we have denoted a
tableau with $\lambda$ boxes in the first row and one box
in each of the next
$\beta$ rows by the symbol $\{\lambda,1^{\beta}\}$.
Now the crucial point about this equation is that the symmetry patterns
of the first term in square brackets on the second line clash with
those of all the terms on the last line. This means that the
first line must vanish by itself. For maximal $f$ the first term
in the square bracket must vanish by itself, which implies
that $\ket{\phi,\{\beta,1^{f}\}}=0$ for this maximal $f$. Then 
induction shows
that they all vanish unless $f=\beta=0$. Thus to find the
cohomology of $Q$ we may use the restricted basis
\bea
&&{\hat F}_0^{J}
{\hat L}_{-\{l_1}
\cdots {\hat L}_{-l_l\}}{b}_{-[m_1}
\cdots {b}_{-m_b]}%
{\hat c}_{-[n_1}
\cdots {\hat c}_{-n_c]}{K}_{-\{p_1}
\cdots {K}_{-p_k\}}\nonumber\\ &&\hskip2in
{\hat \gamma}_{-\{q_1}
\cdots {\hat \gamma}_{-q_\gamma\}}{D}_{-[r_1}
\cdots {D}_{-r_d]}c_0^M{\gamma}_0^N\ket{T}
\label{bigbasis2}
\eea
Now we repeat the argument for the $Lb$ factors (exactly
as in \cite{thorndetailed}) using (\ref{lbqbracket})
which shows that the symmetry pattern
\bea
b+1\Biggl\{\overbrace{\tableau{6 1 1 1 1 1 1}}^{\ell}\Biggr..
\label{tableauiv}
\eea
can be gauged away, after which the requirement that $Q$ annihilate 
$\ket{\phys}$ shows that terms with $Lb$ factors are absent.
One then moves on to the ${\hat c} K$ factors using (\ref{ckqbracket})
and then the ${\hat\gamma} D$ using (\ref{gammadqbracket}).
Thus repeating the same argument four times shows, finally, that the basis 
for calculating the cohomology can be reduced to
\bea
{\hat F}_0^Jc_0^M\gamma_0^N\ket{T}={F}_0^Jc_0^M\gamma_0^N\ket{T}
\eea
which establishes (\ref{zerobasis}) and hence completes the calculation
of the cohomology of $Q$.

\vskip14pt
\noindent\underline{Acknowledgments}: I would
like to thank Edward Witten for a question which inspired
me to add the section on BRST cohomology.
This research was supported in part by the Department
of Energy under Grant No. DE-FG02-97ER-41029.

\end{document}